\documentclass[twocolumn,prl,aps,superscriptaddress,showpacs]{revtex4}
\usepackage{graphicx}
\usepackage{epsf}
\begin{document}
\title{Can Hyperfine Excitation explain the Observed Oscillation-Puzzle
of Nuclear Orbital Electron Capture of Hydrogen-like Ions?}
\author{Nicolas Winckler}
\affiliation{Gesellschaft f{\"u}r Schwerionenforschung (GSI),
Planckstrasse 1, D-64291 Darmstadt, Germany}
\author{Katarzyna Siegie\'n-Iwaniuk}
\affiliation{Soltan Institute for Nuclear Studies, Hoza 69,
PL-00-681 Warsaw, Poland}
\author{Fritz Bosch}
\affiliation{Gesellschaft f{\"u}r Schwerionenforschung (GSI),
Planckstrasse 1, D-64291 Darmstadt, Germany}
\author{Hans Geissel}
\affiliation{Gesellschaft f{\"u}r Schwerionenforschung (GSI),
Planckstrasse 1, D-64291 Darmstadt, Germany}
\author{Yuri Litvinov}\email{Y.Litvinov@gsi.de}
\thanks{On leave from GSI Helmholtzzentrum f\"ur Schwerionenforschung}%
\affiliation{Max-Planck-Institut f\"ur Kernphysik, Saupfercheckweg 1, D-69117 Heidelberg, Germany}%
\author{Zygmunt Patyk}
\affiliation{Soltan Institute for Nuclear Studies, Hoza 69,
PL-00-681 Warsaw, Poland}
\begin{abstract}
Modulated in time orbital electron capture (EC) decays have been observed recently in stored H-like $^{140}$Pr$^{58+}$ and $^{142}$Pm$^{60+}$ ions.
Although, the experimental results are extensively discussed in literature, a firm interpretation has still to be established.
Periodic transitions between the hyperfine states could possible lead to the observed effect.
Both selected nuclides decay to stable daughter nuclei via allowed Gamow-Teller transitions.
Due to the conservation of total angular momentum, the allowed EC decay can only proceed from the hyperfine ground state of parent ions.
In this work we argue that periodic transitions to the excited hyperfine state (sterile) in respect to the allowed EC decay ground state cannot explain the observed decay pattern.
\end{abstract}

\maketitle

Experiments studying electron capture decay of stored highly-charged $^{140}$Pm and ${}^{142}$Pm ions have been performed recently at GSI, Darmstadt.
In these experiments, the highly-charged radioactive ions were produced via fragmentation of about 500~MeV/u $^{152}$Sm projectiles.
The mono-isotopic beams in a selected ionic charge-state were separated in-flight by the fragment separator FRS \cite{FRS} and stored in the storage-cooler ring ESR \cite{ESR}.
In the ESR, the ions were cooled employing first the stochastic \cite{No-NIM} and then the electron cooling \cite{St-NIM}.
The cooled ions were circulating in the ESR with revolution frequencies of about 2~MHz.
Their decay properties have been measured with time-resolved Schottky Mass Spectrometry (SMS)~\cite{{Ra-NPA},{Li-NPA04},{FGM}}.
Each ion at each revolution induced a mirror charge on a pair of capacitive pick-up plates installed inside the ring aperture.
The subsequent fast Fourier transform yielded revolution frequency spectra.
While the frequencies in such spectra reflect the mass-over-charge ratios of the ions \cite{{FGM},{Li-NPA05}},
the areas of the frequency peaks are proportional to the corresponding number of stored ions and to the square of their atomic charge \cite{{FGM},{Bo-IJMS}}.
\begin{table}[b]
\caption{Measured $\beta^+$, EC and atomic-loss decay constants  for fully-ionized,
H-like (in bold), and He-like $^{140}$Pr~\cite{Lit1} and $^{142}$Pm~\cite{Win} ions obtained in MI measurements (see text).
The decay constants are given in the rest frame of the ions.
The total decay constant $\lambda=\lambda_{EC}+\lambda_{\beta^+}+\lambda_{loss}$ is presented in the last column.} \label{lambdas}
\begin{center}
\begin{tabular}{|c|c|c|c|c|}
\hline%
Ion&$\lambda_{\beta^+}$ $[s^{-1}]$&$\lambda_{EC}$ $[s^{-1}]$&$\lambda_{loss}$ $[s^{-1}]$&$\lambda$ $[s^{-1}]$\\
\hline%
\hline%
$^{140}$Pr$^{59+}$ & 0.00158(8) & --- &0.0003(1)&0.0019(1)\\
\hline%
$ {}^{\textbf{140}}\textbf{Pr}^{\textbf{58}+} $ & 0.0016(1) & 0.00219(6) &0.0003(1)&\textbf{0.0041(1)}\\
\hline%
$^{140}$Pr$^{57+}$ & 0.0015(1) & 0.00147(7) &0.0003(1)&0.0033(1)\\
\hline%
\hline%
$^{142}$Pm$^{61+}$ & 0.0123(7) & --- &0.0004(1)&0.0127(7)\\
\hline%
${}^{\textbf{142}}\textbf{Pm}^{\textbf{60}+} $ & 0.0126(3) & 0.0051(1) &0.0004(1)&\textbf{0.0181(3)}\\
\hline%
$^{142}$Pm$^{59+}$ & 0.0139(6) & 0.0036(1) &0.0004(1)&0.0179(6)\\
\hline%
\end{tabular}
\end{center}
\end{table}

Two methods have been developed for the decay studies \cite{Bo-IJMS}.
In the first method, the number of stored parent and daughter ions has been monitored in time.
A few ten to a few hundred parent $^{140}$Pm and ${}^{142}$Pm ions have been stored as fully-ionized, hydrogen- (H-like) or helium-like (He-like) ions \cite{{Lit1},{Win}}.
Constants of the three-body $\beta^+$, $\lambda_{\beta^+}$, and orbital electron capture (EC), $\lambda_{EC}$, decays as well as the atomic-loss constant, $\lambda_{loss}$, have been determined,
where the latter is mainly due to collisions with the rest gas atoms and atomic capture of the electrons in the cooler.
In Table I we summarize the decay constants obtained in Refs. \cite{{Lit1},{Win}}.
In the following we will refer to this method as the many-ion (MI) spectroscopy.
\begin{table}[!t]
\caption{The decay parameters obtained in SI measurements (see text) for $^{140}$Pr$^{58+}$ (upper part) and
$^{142}$Pm$^{60+}$ (lower part) \cite{OSC}.
In case a), the fits of the data points have been done assuming the pure exponential decay
$N'_{EC}(t)=N(0) \cdot \lambda_{EC} \cdot e^{-\lambda t}$ (the quantity $N(0) \cdot \lambda_{EC}$ is constant), where
$\lambda=\lambda_{EC}+\lambda_{\beta^+}+\lambda_{loss}$ and the prime index denotes the time derivative.
In case b), a modulated in time EC decay constant has been assumed
$N'_{EC}(t)=N(0) \cdot \lambda_{EC} \cdot e^{-\lambda t}(1+a \times cos(\omega t+\varphi))$.
The decay constants $\lambda$ for $^{140}$Pr have large uncertainties
due to a short time of  the total observation~\cite{OSC}.}
\label{ta-fits}
\begin{center}
\begin{tabular}{|c|c|c|c|}
\hline%
\multicolumn{4}{|c|}{Fit parameters of $^{140}$Pr$^{58+}$ data} \\
\hline%
Method &$\lambda$ $[s^{-1}]$&$a$&$\omega$ $[s^{-1}]$ \\
\hline%
a)&\textbf{0.0014(10)}&  -    & -     \\
\hline%
b)&\textbf{0.0015(10)}&0.18(3)&0.89(1)\\
\hline%
\multicolumn{4}{|c|}{Fit parameters of $^{142}$Pm$^{60+}$ data}\\
\hline%
Method &$\lambda$ $[s^{-1}]$&$a$&$\omega$ $[s^{-1}]$\\
\hline%
a)&\textbf{0.024(4)}&  -    & -      \\
\hline%
b)&\textbf{0.022(4)}&0.23(4)&0.89(3) \\
\hline%
\end{tabular}
\end{center}
\end{table}

The second method has been applied in Ref. \cite{OSC} and employs the sensitivity of SMS to single stored ions \cite{{Ra-NPA},{Li-NPA04},{FGM}}.
At maximum three parent ions were simultaneously injected into the ESR and their frequencies, that is the masses, were monitored in time.
In EC decay the mass of the ion changes by the corresponding $Q_{EC}$ value while the atomic charge-state is preserved.
Therefore, the EC decays were unambiguously identified by observing sudden changes in the revolution frequency of the ions.
Several thousands EC decays have been measured in this way.
We refer to this method as the single-ion (SI) spectroscopy.
Surprisingly, it has been observed that the number of EC decays per time unit deviates from the expected exponential decay.
The data were described by adding a modulation term $(1+0.2\times {\rm cos}(0.885 \cdot \Delta t))$, where $t$ is the time after the injection of the ions into the ESR,
superimposed on the exponential decay \cite{OSC}. The fit parameters taken from Ref. \cite{OSC} are given in Table II.

The interpretation of the observation above has attracted the attention of many physicists working in different fields.
Quantum beat phenomenon due to the emitted neutrinos, which are flavour - but not mass eigenstates of the weak-interaction hamiltonian,
has been suggested by several authors, see for instance Refs. \cite{{lipkin},{Ivanov1}}.
However, this possibility is strongly disputed, see for example Refs. \cite{{glashow},{Giunti}}.
An alternative explanation has been proposed based on the coupling of the electron and nuclear spins to the rotation in the ring (Thomas precession) \cite{lambiase}.

In this Paper we investigate the hypothesis of periodic transfers between the hyperfine states (e.g. due to interactions with electromagnetic fields or with the cooling system)
of H-like $^{140}$Pr$^{58+}$ and $^{142}$Pm$^{60+}$ ions.
Due to conservation of the total angular momentum, the upper hyperfine state does not decay by allowed EC decay \cite{Lit1} which can cause the observed modulations.

A detailed theoretical description of the EC decay rates in H-like and He-like ions has been performed in Refs. \cite{{Pa-PRC},{Ki-PRC}}.
There, it has been proven that the ratio of these rates  equal to $\lambda_{EC}^{\rm H-like}\approx3/2 \cdot \lambda_{EC}^{\rm He-like}$ to a few percent.
Based on this fact, we will show that the measured  decay constants in Refs. \cite{{Lit1},{Win}}  (see Table I) cannot accomodate the hypothesis of periodic transfers between the hyperfine states.

Moreover, we discuss the possibility of observing the modulated decays also in MI experiments,
which, if successful, would be of a great advantage due to the small statistics presently achievable in SI measurements.

Both investigated nuclei $^{140}$Pr and $^{142}$Pm  have spin $I=1^+$ and decay by allowed Gamow-Teller transition to the ground state of stable daughter nuclei with spin $0^+$.
The parent H-like ions can therefore have two hyperfine states $|-\rangle$ and $|+\rangle$
each with total angular momenta $F^-=I-1/2$ and $F^+=I+1/2$, respectively.
The order of the hyperfine states depends on the sign of the corresponding nuclear magnetic moment.
For the magnetic moment parallel to the nuclear spin the hyperfine ground state has spin $F^-$ and in the opposite case it is $F^+$.
However, in the allowed Gamow-Teller EC transitions $I\rightarrow I\pm1$, due to conservation of the total angular momentum,  only states can decay that have spins $F^\pm$.

The energy splitting $\delta E$ between the two states $|-\rangle$ and $|+\rangle$ can be estimated by using \cite{Shabaev1}:
\begin{eqnarray}\label{4c}
\delta E&=&\frac{4}{3} \alpha (\alpha Z)^3 \frac{\mu}{\mu_N}
\frac{m}{m_p}\frac{2I+1}{2I} A( \alpha Z) mc^2,
\end{eqnarray}
where $m$, $m_p$, $\mu$, $\mu_N$, and  $\alpha$ are the electron mass, proton mass, nuclear magnetic moment, nuclear magneton, and the fine-structure constant, respectively.
The relativistic factor
$A( \alpha Z)$ is defined as follows
\begin{eqnarray}\label{4d}
A( \alpha Z) = \frac{1}{(2 \sqrt{1-(\alpha Z)^2}-1)\sqrt{1-(\alpha
Z)^2}}.
\end{eqnarray}
We note, that the corrections for the nuclear charge distribution, the nuclear
magnetization distribution and the QED effects are not included in Eq.~(\ref{4c}).

\begin{table}
\caption{Hyperfine splitting parameters. The ion and the nuclear transition are given in the first two columns.
Nuclear magnetic moments used in this work are given in the second column.
The hyperfine splitting $\delta E$ and the relaxation decay constants $\lambda_{hf}$ are given in the third and fourth columns, respectively.
The excitation constant $b$ (see text) is given in the last column.}
\label{aabb}
\begin{tabular}{|c|c|c|c|c|c|}
\hline
Ion  & Transition  & $\mu / \mu_N$ &$\delta E$ [eV] & $\lambda_{hf}$ $[s^{-1}]$ & $b$ $[s^{-1}]$  \\
\hline
$^{140}$Pr  & $1^+\rightarrow 0^+$                     & +2.5    & 1.26  & 38.2    & 7.8 \\
$^{142}$Pm  & $1^+\rightarrow 0^+$                     & +2.5    & 1.12  & 26.2    & 5.4 \\
\hline
\end{tabular}
\end{table}

The probability for a spontaneous transition from the excited hyperfine state
$|\pm\rangle$ with angular momentum $F^\pm$  to the ground state $|\mp\rangle$ is expressed by a simple formula~\cite{{Sobelman},{Shabaev2}}
\begin{eqnarray}\label{4d}
\lambda_{hf}= \frac{4 \alpha}{3}  {\delta E}^3
\frac{1}{\hbar m^2 c^4} \frac{2F^\mp+1}{2I+1}.
\end{eqnarray}
The calculated energy splitting $\delta E$ and the decay probability $\lambda_{hf}$  are presented for H-like $^{140}$Pr and $^{142}$Pm ions in Table~\ref{aabb}.
The magnetic moment of $^{140}$Pr has been estimated in Ref.~\cite{Borzov}.
The $^{142}$Pm nucleus has a similar structure, therefore we applied the same magnetic moment values as in the case of $^{140}$Pr.
Note, the hyperfine decay constant  $\lambda_{hf}$ depends on the ninth power of the atomic number $Z$.
In our case, the ground state of H-like ions  $|-\rangle$ has the spin $F^-$ and can decay via EC and $\beta^{+}$ decay.
However, the excited state $|+\rangle$ can only decay by $\beta^+$ decay. We note, that ions in both hyperfine states can be lost due to atomic interactions.
We will denote the ions as {\it active} or {\it non-active} to distinguish the ions which can or cannot decay by the allowed EC decay, respectively.
The number of active ions (A) in the state $|-\rangle$ is $N_A(t)$ and the number of non-active (N) nuclei in the state $|+\rangle$ is $N_N(t)$

\begin{figure}[h]
\begin{center}\includegraphics[width=0.85\linewidth]{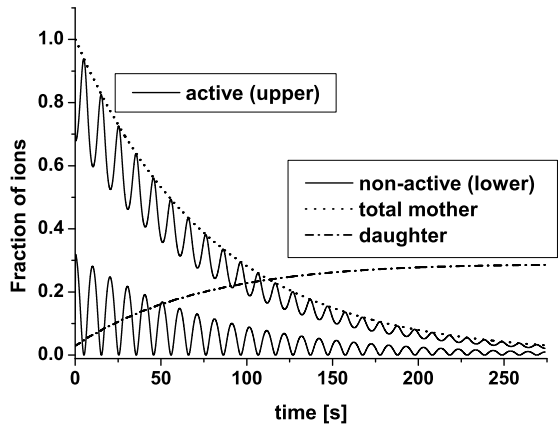}\end{center}
\caption{\label{fig1} Fractions of EC active mother  nuclei (upper solid), non-active mother
nuclei (lower solid) and  the total number of mother nuclei (dot) and fraction of daughter nuclei (dashed-dotted)
as a function of laboratory time. Note,  the total numbers of mother and daughter
nuclei almost do not oscillate in time. }
\end{figure}

We simulate the time-modulation of the number of EC decaying ions by introducing a
simple mechanism: nuclei in the A-state $|-\rangle$ are periodically excited with a probability $b\times\{1+{\rm cos}(\omega  t+\varphi)\}$ to the N-state $|+\rangle$.
The excited state $|+\rangle$ decays  spontaneously to the ground state $|-\rangle$ with decay constant $\lambda_{hf}$ (See Fig.~\ref{fig1}).
Experimentally, such periodic excitations could be due to motion in the electromagnetic fields of the ESR or due to spin-flip reactions in the cooler.
The relevant hyperfine splitting parameters are summarized in Table~\ref{aabb}.
The modulation of the EC decay constant is described by the set of four parameters $\{\lambda_{hf}, b, \omega, \varphi\}$.
For the sake of simplicity we put $\varphi=0$.
We can write three coupled differential equations for the number of parent A-ions $N_A(t)$, N-ions  $N_N(t)$, and for the number of daughter nuclei $N_D(t)$:
\begin{eqnarray}\label{2a}
N'_A(t)&=&-(\lambda_{EC}+\lambda_{loss}+\lambda_{\beta_+}) N_A(t)\nonumber\\
&+&\lambda_{hf} N_N(t)-b\times\{1+{\rm cos}(\omega t)\}N_A(t)\label{1},\\\label{2b}
N'_N(t)&=&-(\lambda_{loss}+\lambda_{\beta_+}) N_N(t)\nonumber\\
&-&\lambda_{hf} N_N(t)+b\times\{1+{\rm cos}(\omega t)\}N_A(t)\label{2},\\\label{2c}
N'_D(t)&=& -\lambda_{loss}N_D(t) +\lambda_{EC}N_A(t),
\end{eqnarray}
where the prime index denotes the time derivative. These equations have been solved numerically using the Euler method. A time step of 0.00056 s has been chosen.
\begin{figure}[h]
\begin{center}\includegraphics[width=0.85\linewidth]{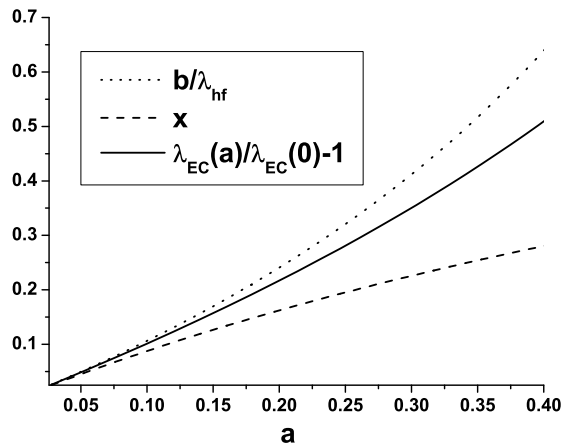}\end{center}
\caption{\label{fig2}  The ratio b/$\lambda_{hf}$ (dotted), the renormalization factor
$\lambda_{EC}(a)/\lambda_{EC}(0)$ (solid) and the average occupancy
of the excited hyperfine state (dashed) as a function of the oscillation
amplitude $a$. (The meaning of the vertical axis is given in the legend.)}
\end{figure}
In the beginning we establish an approximate relation between the constants $\lambda_{hf}$ and $b$.
Let us define
\begin{equation}\label{5b}
x=\frac{<N_N(t)>}{<(N_N(t)+N_A(t))>}
\end{equation}
which is the ratio of the average number of N-nuclei to the total number of parent ions.
An average number of ions decaying from the N-state to the A-state ($N\rightarrow A$) approximately equals to  $x\lambda_{hf}$ and
it should be equal to the number of ions  $b(1-x)$ being excited from the A-state to the N-state ($A\rightarrow N$)
\begin{equation}\label{5a}
x \lambda_{hf} \approx b(1-x)\Rightarrow x\approx \frac{b/\lambda_{hf}}{1+b/\lambda_{hf}}.
\end{equation}
For a fixed ratio $x$,  the parameters $\lambda_{hf}$ and $b$ are free.
Using Eq. \ref{5b} the ratio $x$ can be connected to the modulation amplitude $a$: $x\approx a/(1+a)$.
Comparing this relation with Eq. \ref{5a} we find that $a\approx b/\lambda_{hf}$
For the observed $a\approx 0.2$~\cite{OSC} we obtain the ratio $x \approx 0.17$.
The numerical solutions of Eqs.~ \ref{2a}-\ref{2c} are plotted in Fig. \ref{fig1}.
We observe that, if we introduce the modulation parameters derived from SI experiments,
we are not able to describe the decay curves measured with MI method (for the fit parameters see Tables~\ref{lambdas}-\ref{ta-fits}).
The reason for this is simple: effectively only the $1-x$ fraction of parent nuclei is active.
Thus, the effective decay constant $\lambda\approx(1-x)\lambda_{EC}+\lambda_{\beta}+\lambda_{loss}$ is smaller since some of the ions are   in "sterile" N-state.
Therefore, if the periodic transfer between the hyperfine states occurs in the MI experiments, the decay constants obtained in H-like ions shall be readjusted.
Due to the absence of hyperfine splitting, no such adjustment is needed in the case of He-like ions.
The measured ratio of decay constants $\lambda_{EC}^{H-like}/\lambda_{EC}^{He-like}$ has to be corrected approximately by $1/(1-x)=1.20$.
The numerical solutions of Eqs. \ref{2a}-\ref{2c} fitted to the experimental data ~\cite{Lit1}
gives the correction factor of 1.23.
However, the readjusted experimental ratios for $^{142}$Pm and ${}^{140}$Pr ions are 1.77(7) and 1.83(10) which disagree, respectively, by about 4 and 3 standard deviations to the theoretical ratio of 1.5.

The numerical solutions of Eqs. \ref{2a}-\ref{2c} for a fixed value of the oscillation amplitude $a$
were fitted to the MI experimental mother and daughter populations~\cite{Lit1}. As a result we
obtained the ratio $b/\lambda_{hf}$, the average occupancy of the excited hyperfine state $x$ and the
EC decay constant $\lambda_{EC}(a)$. The results as a function of the amplitude a are plotted
in Fig.~\ref{fig2}.

Let us analyze a different model and assume that indeed the electron capture probability
varies in time as $1+0.2 \ {\rm  cos}(0.89 \  \Delta t+\varphi)$,
where $\Delta t$ is the time interval from the creation until the decay of the ion.
The origin of such modulations is still intensively discussed.
Experiments with implanted neutral ${}^{142}$Pm and ${}^{180}$Re atoms, performed in Berkeley~\cite{Vetter} and in Garching~\cite{Bosch2},
respectively, have not  observed time-modulations of EC decays.
%

The mathematical model with the time-modulated EC constant $\lambda_{EC}$ has the following form:
\begin{eqnarray}\label{12a}
dN_M(t)&=&-\lambda_{EC}\{1+a\ cos(\omega t+\varphi)\}N_M(t)dt\nonumber\\
&-&(\lambda_{loss}+\lambda_{\beta_+}) N_M(t)dt,\\
dN_D(t)&=& \ \lambda_{EC}\{1+a\ cos(\omega t+\varphi)\}N_M(t)dt\nonumber\\
&-& \lambda_{loss}N_D(t)dt,\label{12b}.
\end{eqnarray}
The decay constants can be taken from Table~\ref{lambdas} and the corresponding time-modulated parameters $a$ and $\omega$ can be taken from Table~\ref{ta-fits}.
The solution of Eq.~\ref{12a} can be written as:
\begin{eqnarray}\label{13a}\nonumber
N_M(t)=N_M(0) e^{-\lambda_{loss}t-\lambda_{\beta_+}t-\lambda_{EC}(t-a \ sin(\omega t+\varphi)/\omega)}.\nonumber
\end{eqnarray}\nonumber

The Eq.~\ref{12b} has been integrated analytically and also checked numerically.
The obtained results
show that the assumption of a time-dependent decay probability observed in SI measurements is in perfect agreement with the decay and growth curves extracted
from the experimental data of MI experiments.

However, we can construct a different approach and  assume that all decay constants are time-modulated or, what is mathematically equivalent, that the time scale fluctuates. The numbers of mother $N_M(t)$ and
daughter $N_D(t)$ nuclei are connected by the two following equations:

\begin{eqnarray}\label{22a}
dN_M(t)&=&-dt\{1+a \ cos(\omega t+\varphi)\}\times\nonumber\\
&&(\lambda_{EC}+\lambda_{loss}+\lambda_{\beta_+}) N_M(t),\\
dN_D(t)&=& dt\{1+a \  cos(\omega t+\varphi)\}\times \nonumber\\
 &&(\lambda_{EC}N_M(t)- \lambda_{loss}N_D(t))\label{22b}.
\end{eqnarray}

Eqs \ref{22a}-\ref{22b} can be easily solved by introducing a fluctuating time scale
$t'$ connected with  time t by the relation $t'=t-a \ sin(\omega
t+\varphi)/\omega$. Thus, the numbers of mother and  EC daughter ions can be
expressed by analytic  functions of time $t'$ (see e.g. ~\cite{Lit1}) . For large t (compared
with $1/\omega)$ the ratio of both times is approaching one. We see that experimental
populations in MI can be fitted with almost the  same parameters for $a \neq0$
and for $a=0$.

We have shown that if the modulation appears only for H-like ions then the ratio $\lambda_{EC}^{H-like}/\lambda_{EC}^{He-like}\approx 3/2$ ~\cite{{Pa-PRC},{Ki-PRC}} has to be multiplied by 1.23.
Such an increased value is very improbable from the view of the present experimental and theoretical data.
The $\lambda_{EC}^{H-like}/\lambda_{EC}^{He-like}$ ratios from MI experiments are in excellent agreement with the factor $3/2$.
This disagreement could be resolved if the described modulation appears in He-like ions as well.
However, there are no hyperfine states in He-like ions and the discussed mechanism is physically disabled.
We emphasize, that in order to describe consistently the present data of MI and SI experiments, the modulation phenomenon with similar amplitude has also to occur in He-like ions.
Therefore, it is indispensable to perform SI experiments on He-like ${}^{142}$Pm$^{59+}$ or ${}^{140}$Pr$^{57+}$ ions.
Furthermore, the ratio $\lambda_{EC}^{H-like}/\lambda_{EC}^{He-like}$ turned out to be a sensitive probe and more accurate measurements are, therefore, required.

We have investigated the consistency of the measured data in respect to the modulated EC decay constant.
We have shown that the 20~\% modulation amplitude determined from SI experiments translates into a tiny, much less than a percent, modulation amplitude of decay curves measured with the MI spectroscopy.
Such accuracy is presently out of reach for the MI experiments.

\end{document}